\documentclass[12pt,a4paper]{article}

\usepackage{amsthm,amsmath,natbib,amssymb,amsfonts,bm, mathtools}
\usepackage{graphicx}
\usepackage{placeins}
\usepackage{epstopdf}
\usepackage{secdot}

\usepackage{sectsty} 
\sectionfont{\fontsize{12}{16}\selectfont} 
\subsectionfont{\fontsize{12}{16}\selectfont} 

\def\hang{\hangindent\parindent}
\def\rf{\par\noindent\hang}

\newtheorem*{theorem*}{Theorem}

\theoremstyle{definition}

\theoremstyle{remark}

\renewenvironment{proof}{{\bfseries Proof.}}

\DeclareMathAlphabet{\mathpzc}{OT1}{pzc}{m}{it}

\newcommand{\SSB}{\text{\small SSB}}
\newcommand{\SSW}{\text{\small SSW}}

\def\hang{\hangindent\parindent}
\def\rf{\par\noindent\hang}

\begin{document}

\baselineskip=20pt

\begin{center}
{\bf \Large Confidence Intervals that Utilize Uncertain Prior Information about Exogeneity in Panel Data}
\end{center}

\medskip

\begin{center}
{\bf \large Paul Kabaila$^*$ and Rheanna Mainzer$^{**}$}
\end{center}


\begin{center}
{\large
{\sl La Trobe University}
}
\end{center}

\bigskip

\begin{center}
	{\bf ABSTRACT}
\end{center}

\noindent Consider panel data modelled by a linear random intercept model that includes a time-varying covariate. 
Suppose that we have uncertain prior information that this covariate is exogenous. 
We present a new confidence interval for the slope parameter that utilizes this uncertain prior information. This interval has minimum coverage probability very close to its nominal coverage. 
Let the scaled expected length of this new confidence interval be its expected length divided by the expected length of the 
confidence interval, with the same minimum coverage, constructed using the {\sl fixed effects model}. This new interval has scaled expected length
that (a) is substantially less than 1 when the 
prior information is correct,
(b) has a maximum value
that is not too much larger than 1
and (c) is close to 1 
when the data strongly contradict the prior information. 
We illustrate the properties of this new interval using 
an airfare data set.

\vspace{6cm}

\noindent * Corresponding author. Department of Mathematics and Statistics, La Trobe University, Victoria 3086, Australia. e-mail: P.Kabaila@latrobe.edu.au

\bigskip

\noindent ** Department of Mathematics and Statistics, La Trobe University, Victoria 3086, Australia. e-mail: rmmainzer@students.latrobe.edu.au

\newpage


\section{INTRODUCTION}


We consider panel data modelled by a linear random intercept model that includes a time-varying covariate. Irrespective of whether or not this covariate is exogenous, 
valid inference results from the {\sl fixed effects model}. However, if this covariate is exogenous then valid and more efficient inference results from the {\sl random effects model}. This greater efficiency motivates the use of this model, whenever
it is appropriate, for inference.

The Hausman (1978) pretest is commonly used in practice to decide whether or not 
subsequent inference will be based on the assumption that 
the time-varying covariate is exogenous.  If the null hypothesis that this covariate is exogenous is accepted then the {\sl random effects model} is used for subsequent inference; otherwise the {\sl fixed effects model} is used (see e.g. Hastings,  2004, Papatheodorou and Lei, 2006, Choe, 2008, and Jackowicz, Kowalewski and Kozlowki, 2013.). 
A somewhat similar approach is to report the inference resulting from both of these models, and to state the preferred model based on the outcome of the Hausman pretest (see e.g. Smith, Smith and Verner, 2006, and Stanca, 2006).
This pretest is incorporated in popular software packages such as \texttt{eViews}, \texttt{R}, \texttt{SAS} and \texttt{Stata}.

 Guggenberger (2010) and Kabaila, Mainzer and Farchione (2015, 2017) show that the effect of model selection using the Hausman pretest in this way 
 is very damaging to hypothesis tests and confidence intervals for the slope parameter.
 In particular, the confidence interval constructed
 using the model selected by the Hausman pretest has minimum coverage probability that is typically far below the nominal coverage. 
 This is an example of a widespread problem with confidence intervals 
 constructed after preliminary data-based model selection (see e.g. Kabaila, 2009, and Leeb and P\"otscher, 2005).
As in Kabaila \textit{et al} (2017), throughout the present paper we carry out inference conditional
on the observed values of the time-varying covariate. The very important advantages of this conditional inference are described in the introduction to Kabaila \textit{et al} (2017). These advantages include the fact that this inference is valid irrespective of how the values of the covariate are generated.  In the econometric literature, an early recognition of this advantage is provided by Koopmans (1937, pp 29 and 30).
In the present paper, the inference of 
interest is a 
confidence interval for the slope parameter.

As insightfully pointed out by Leamer (1978, chapter 5), preliminary data-based model selection (such as a Hausman pretest) may be motivated by the desire to utilize uncertain prior information in statistical inference.  He goes even further when he states,
on p.123, that ``The mining of data that is common among non-experimental scientists constitutes prima facie evidence of the existence of prior information''.
Leamer uses a Bayesian approach to utilize uncertain prior information 
in statistical inference. An alternative is to use a frequentist approach
to utilize the uncertain prior information. 
The utilization of
uncertain prior information in frequentist inference has a 
distinguished history. Hodges and Lehmann (1952), Pratt(1961), Stein (1962), Cohen (1972),  Bickel (1984), Kempthorne (1983, 1987, 1988), Casella and Hwang (1983, 1987), 
Goutis and Casella (1991), Tseng and Brown (1997) and  Efron (2006) 
are included in this history, which is briefly reviewed by Kabaila (2009, Sections 8 and 9).

In the present paper we use the frequentist approach and suppose that there is, indeed,   
uncertain prior information that the covariate is exogenous.
If this prior information was certain then we would simply use the 
{\sl random effects model} to construct a confidence interval with desired confidence coefficient $1 - \alpha$ for the slope parameter. Recall that the confidence 
coefficient of a confidence interval is defined to be the infimum over the parameter
space of the coverage probability of this interval.
However, since we assume only {\sl uncertain} prior information, we seek a 
confidence interval with the desired confidence coefficient and an expected length that (a) is relatively small  
when the prior information is correct and (b) is not too large 
when the prior information happens to be incorrect.

We show how the uncertain prior information that the covariate is exogenous
can be utilized to construct a new confidence interval for the slope parameter with the following attractive properties. This confidence interval has a confidence coefficient that is very close to the desired value $1 - \alpha$. 
Define the scaled expected length of this new confidence interval to be the expected length of this interval divided by the expected length of the 
confidence interval, with the same minimum coverage, that is constructed using the {\sl fixed effects model}. This new confidence interval has scaled expected length
that (a) is substantially less than 1 when the prior information is correct and (b) has maximum (over the parameter space) that is not too much larger than 1. In addition, 
the new confidence interval reverts to the confidence interval,  with confidence coefficient $1 - \alpha$ and constructed using the {\sl fixed effects model},
when the data strongly contradict the prior information.
Unlike the endpoints of the confidence interval constructed using a model chosen by a Hausman pretest, which are discontinuous functions of the data, the endpoints of the new confidence interval are smooth functions of the data.

Consider, as an example, the airfare data set provided by Wooldridge (2013).  We are interested in the regression of the response \textsl{lfare}, which is the logarithm of the fare, on the covariate  \textsl{concen}, which is the market share of the largest carrier. This is panel data with the response \textsl{lfare} measured for each of $N = 1149$ city-pair markets over $T = 4$ time points.
Using an analysis conditional on the observed values of this covariate,
 Kabaila \textit{et al} (2017) found that the confidence coefficient of the confidence interval, with nominal coverage 0.95, that results from a Hausman pretest is approximately 0.19, which is much lower than the nominal coverage.  This confidence interval also has a scaled expected length that exceeds 1 throughout the parameter space.

\begin{figure}[h]
\centering
\includegraphics[scale = 0.7]{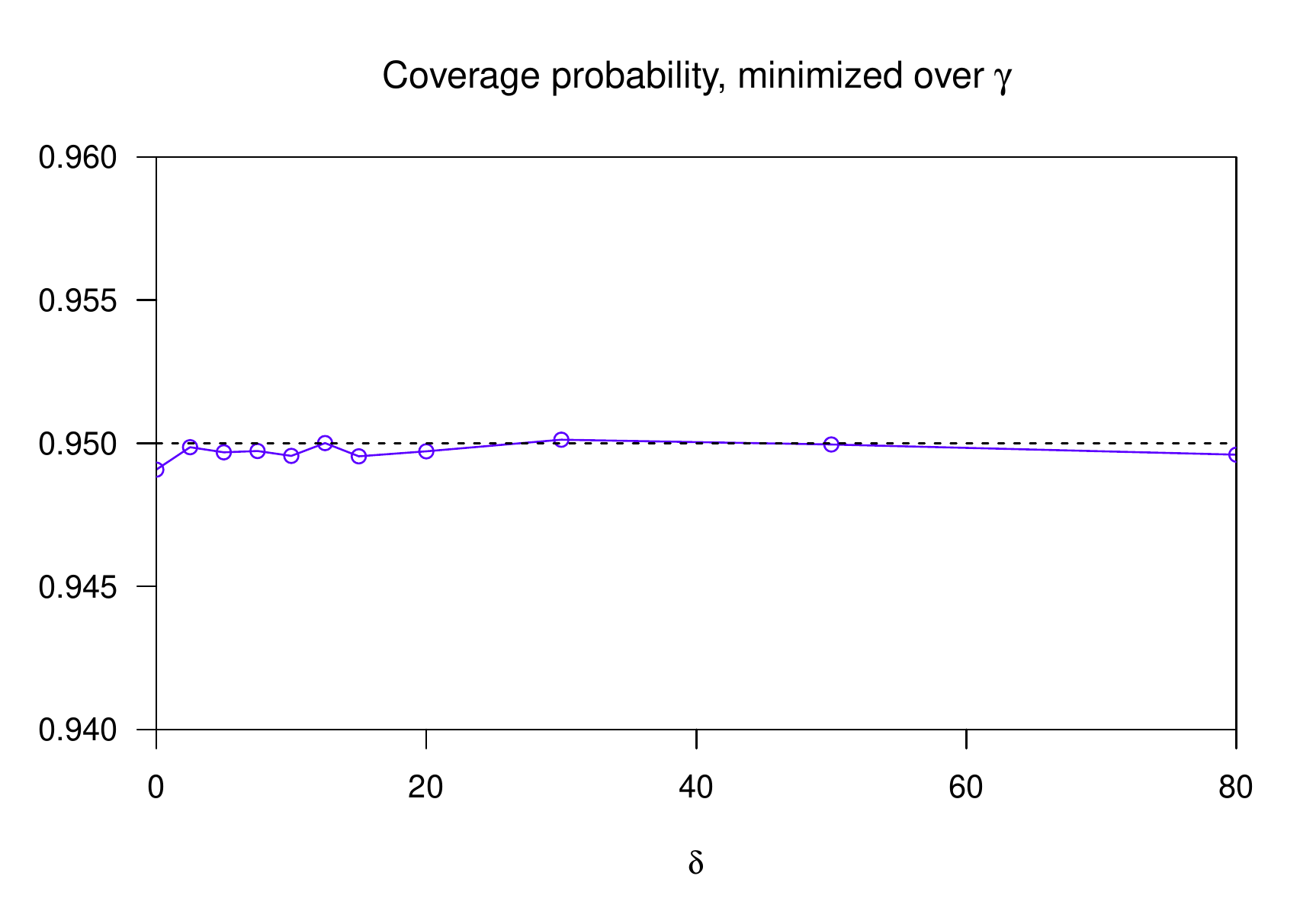}
\vspace{-0.5cm}
\caption{Graph of the coverage probability of the new confidence interval, minimized over $\gamma$, as a function of $\delta$ for the airfare data. Here 
	$N = 1149$, $T = 4$ and	$1 - \alpha = 0.95$. The confidence coefficient is estimated to be 0.9493.
}
\label{cmin_airfare}
\end{figure}

For the linear random intercept model that we consider, the coverage probabilty of the new confidence interval for the slope parameter is a function of two unknown parameters:
 $\delta$, which is the ratio (variance of random effect)/(variance of random error),
 and $\gamma$, a parameter that takes a non-zero value unless the covariate is exogenous. In other words, the covariate is exogenous when $\gamma = 0$.
For this airfare data, the new confidence interval, with desired confidence coefficient
0.95, is found to have confidence coefficient that is approximately 0.9493. 
The coverage probability, minimized over $\gamma$, of this new confidence interval for the airfare data is graphed as a function of $\delta$ in Figure \ref{cmin_airfare}.
%
%

\FloatBarrier

\begin{figure}[h]
	\includegraphics[scale = 0.8]{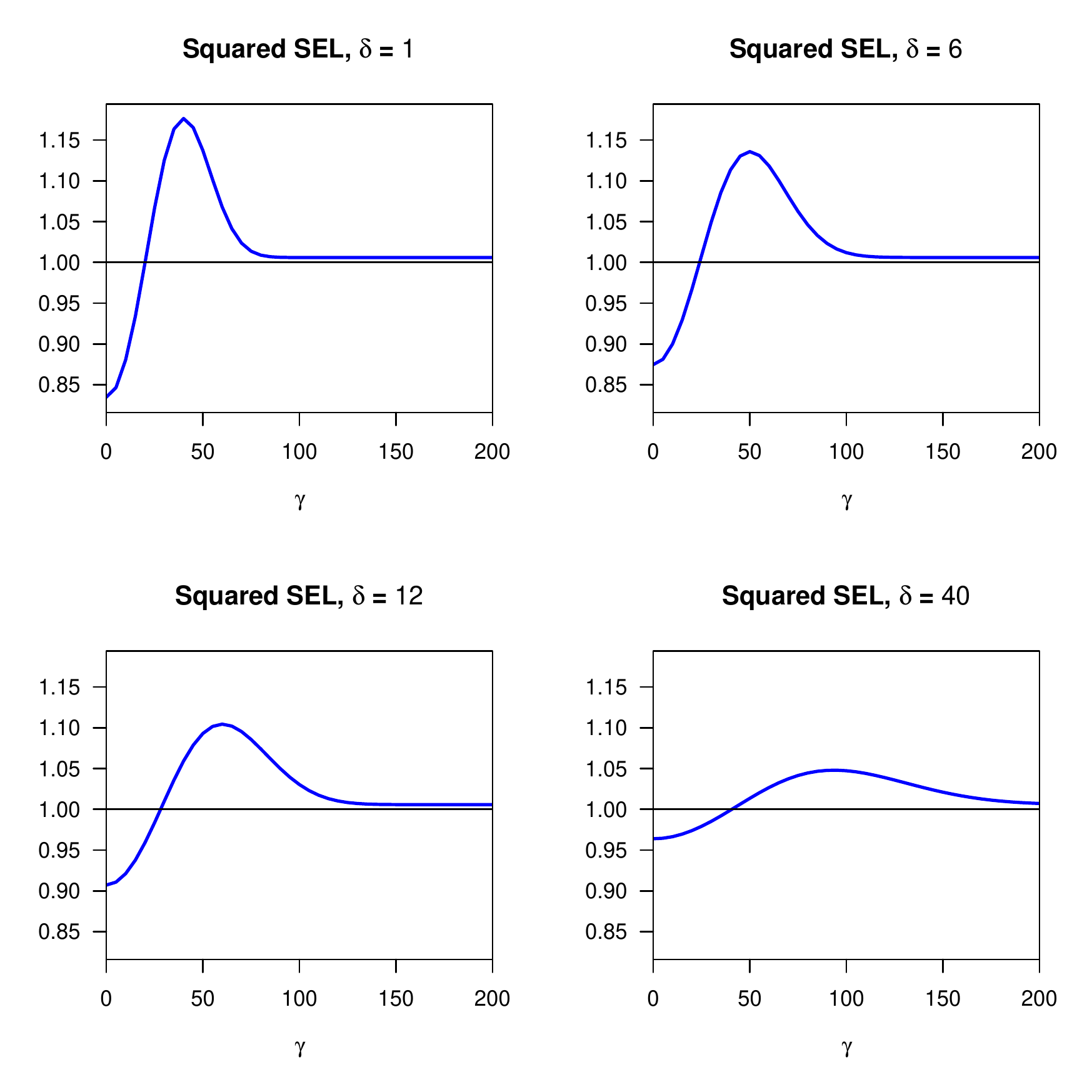}
     \vspace{-0.5cm}
	\caption{Graphs of the squared scaled expected length of the new confidence interval, as a function of $\gamma$ for given $\delta$ for the airfare data. Here 
		$N = 1149$, $T = 4$ and	$1 - \alpha = 0.95$.}
	\label{fig:SEL_airfare}
\end{figure}

Figure \ref{fig:SEL_airfare} presents graphs of the squared scaled expected length of the new confidence interval for the slope parameter, with desired confidence coefficient
0.95, as a function of $\gamma$, for $\delta \in \{1, 6, 12, 40\}$, for the airfare data. The squared scaled expected length is directly related to the relative sample sizes required for the new confidence interval and the 
confidence interval, with the same confidence coefficient and constructed using the {\sl fixed effects model}, to have the same expected length (cf Bickel and Doksum, 1977, p. 137). The graphs of the squared scaled expected lengths were found to be very close to being even functions of $\gamma$ and so Figure 2 presents these graphs only for $\gamma \ge 0$.
How far the squared scaled expected length is below 1 when $\gamma = 0$ (i.e. when the prior information that the covariate is exogenous is correct) depends on the unknown parameter $\delta$. The estimate of $\delta$ for the airfare data is 12.774, suggesting that the graph of the squared scaled expected length of the new confidence interval will be similar to the left-hand lower panel of Figure 2.


\FloatBarrier

 A description of the simulation methods used to prepare Figures 1 and 2 is provided in Section 3. In the next section we describe the correlated random effects model and the new confidence interval which is obtained using the following two steps.
In the first step, we suppose that $\delta$ and the random error variance are known
and construct a confidence interval for the slope parameter that utilizes the uncertain prior information that the covariate is exogenous using the method of Kabaila and Giri (2009). This method has since been used in other contexts, including the construction of optimized Stein-type confidence sets for the multivariate normal mean (Abeysekera and Kabaila, 2017).  In the second step, we replace these parameter values by estimates. In other words, we use the plug-in principle (Efron, 1998, Section 5).

\section{THE CORRELATED RANDOM EFFECTS MODEL AND THE NEW CONFIDENCE INTERVAL}
\label{sec_model}

We consider a model for panel data, for which $i$ denotes the individual, household or firm etc. ($i = 1, \dots, N$) and $t$ denotes time ($t = 1, \dots, T$).
Let $y_{it}$ and $x_{it}$ denote the response variable and the time-varying covariate, respectively, for the $i$'th individual at the $t$'th time. Our analysis is conditional on $x = (x_{11}, \dots, x_{1T}, \dots, x_{N1}, \dots, x_{NT})$.  Suppose that
\begin{equation}
\label{CRE_model}
y_{it} = a + b \, x_{it} + \xi \overline{x}_i + \eta_i + \varepsilon_{it}
\end{equation}
for $i = 1, \dots, N$ and $t=1, \dots, T$, where $\overline{x}_i = T^{-1} \sum_{t=1}^T x_{it}$.  We assume that the $\eta_i$'s and the $\varepsilon_{it}$'s are independent, with the $\eta_i$'s independent and identically distributed (i.i.d.) $N(0, \sigma^2_{\eta})$ and the $\varepsilon_{it}$'s are i.i.d. $N(0, \sigma^2_{\varepsilon})$.  
Let $\delta = \sigma_{\eta}^2 / \sigma_{\varepsilon}^2$. 
The $\varepsilon_{it}$'s and the $\eta_i$'s are unobserved.  
This is the correlated random effects model described, for example, by 
Wooldridge (2013, p.497).
If $\xi = 0$ then the $x_{it}$'s are exogenous. Thus, we call $\xi$ 
a non-exogeneity parameter.
Our aim is to find a confidence interval for the slope parameter $b$ 
with the desired minimum coverage probability and expected length that (a) is relatively small when the prior information is correct and (b) is not too large when the prior information happens to be incorrect.

The new confidence interval is constructed using the following two steps.
In the first step we suppose that $(\sigma_{\varepsilon}, \delta)$ is known
and note that, after the standard transformation described in Appendix A, the model \eqref{CRE_model}
becomes a linear regression model with i.i.d. normal random errors with known 
variance. The method of Kabaila and Giri (2009) is then adapted to this 
particular case (as described in Appendix A) to find a confidence interval that has
the desired coverage and expected length properties.
Of course, this confidence interval is determined by  $(\sigma_{\varepsilon}, \delta)$.  
The second step is to replace  the value of $(\sigma_{\varepsilon}, \delta)$ used in the construction of this confidence interval by the estimator 
 $(\widehat{\sigma}_{\varepsilon}, \widehat{\delta})$ 
described in Subsection \ref{sigma_delta_unknown}.

\subsection{First step: confidence interval construction assuming that $\boldsymbol{(\sigma_{\varepsilon}, \delta)}$ is known}
\label{subsec_CI_sig_del_known}

The details for the first step are as follows. Suppose that 
 $(\sigma_{\varepsilon}, \delta)$ is known.
Adding and subtracting $b \, \overline{x}_i$ to the right hand side of \eqref{CRE_model} gives
\begin{align}
y_{it} 
= a + b_W (x_{it} - \overline{x}_i) + b_B \, \overline{x}_i + \eta_i + \varepsilon_{it}, \label{CRE_model2_l2}
\end{align}
where $b_W = b$ and $b_B = b + \xi$.
Let $\widehat{b}_W$ and $\widehat{b}_B$ denote the GLS estimators of $b_W$ and $b_B$, respectively, obtained using this model.  
Note that $\widehat{b}_W$ is the estimator of $b$ obtained using the {\sl fixed effects model}.
Now let 
\begin{equation*}
h(\sigma_{\varepsilon}, \delta) = \dfrac{\widehat{b}_W - \widehat{b}_B}
{\left(\text{Var}(\widehat{b}_W \, | \, x) 
+ \text{Var}(\widehat{b}_B \, | \, x)\right)^{1/2}},
\end{equation*}
where $\text{Var}(\widehat{b}_W \, | \, x)$
and $\text{Var}(\widehat{b}_B \, | \, x)$ denote the variances of 
$\widehat{b}_W$ and $\widehat{b}_B$, respectively,
conditional on $x$.
Note that $h^2(\sigma_{\varepsilon}, \delta)$ is the test statistic for
a Hausman test of the null hypothesis $\xi = 0$ 
against the alternative hypothesis $\xi \ne 0$.

Let $\SSB = \sum_{i=1}^N (\overline{x}_i - \overline{x})^2$ and
$\SSW = \sum_{i=1}^N \sum_{j=1}^T (x_{it} - \overline{x}_i)^2$
It follows from the proof of Lemma 1 of Kabaila \textit{et al} (2017) that
\begin{equation*}
\text{Var}(\widehat{b}_W \, | \, x) 
= \frac{\sigma^2_{\varepsilon}}{\SSW} 
\quad \text{and} \quad
\text{Var}(\widehat{b}_B \, | \, x) 
= \frac{\sigma^2_{\varepsilon} (\delta + T^{-1})}{\SSB}.
\end{equation*}
We see from this that our assumption that $(\sigma_{\varepsilon}, \delta)$ is known
is needed to find $h(\sigma_{\varepsilon}, \delta)$.
The confidence
interval for $b$, based on the {\sl fixed effects model}
and with coverage probability $c$, is
\begin{equation*}
L(\sigma_{\varepsilon}, c) 
= \left[
\widehat{b}_W - z_{(c+1)/2 } \, \left(\sigma_{\varepsilon}^2 / \SSW \right)^{1/2}, \,
\widehat{b}_W + z_{(c+1)/2 } \, \left(\sigma_{\varepsilon}^2 / \SSW\right)^{1/2}
\right],
\end{equation*}
where $z_p = \Phi^{-1}(p)$ and $\Phi$ denotes the $N(0, 1)$ cdf.

As shown in Appendix A,
after a standard transformation, the model
 \eqref{CRE_model2_l2}
becomes a linear regression model with i.i.d. normal random errors with known 
variance. 
Let $[m \pm w]$ denote the interval $[m - w, m + w]$ ($w \ge 0$).
Using the method of Kabaila and Giri (2009) as described in Appendix A, we are led to a confidence interval 
of the form
\begin{equation*}
J \big(f_o, f_e; \sigma_{\varepsilon}, \delta \big) 
= \left[ 
\widehat{b}_W + \left(\sigma_{\varepsilon}^2 / \SSW \right)^{1/2} f_o\big(h(\sigma_{\varepsilon}, \delta)\big) 
\pm \left(\sigma_{\varepsilon}^2 / \SSW \right)^{1/2} f_e\big(h(\sigma_{\varepsilon}, \delta)\big) \right],
\end{equation*}
where $f_o: \mathbb{R} \rightarrow \mathbb{R}$ is an odd continuous function
and $f_e: \mathbb{R} \rightarrow [0, \infty)$ is an even continuous function.
Obviously, $f_o(0) = 0$.
In addition, we require that $f_o(x) = 0$ and $f_e(x) = z_{1-\alpha/2}$ for $|x| \ge d$,
where $d$ is a (sufficiently large) specified positive number. 
 By construction,  $J(f_o, f_e; \sigma_{\varepsilon}, \delta)$
reverts to the confidence interval,  with confidence coefficient $1 - \alpha$ and constructed using the {\sl fixed effects model},
when $|h| \ge d$ i.e. when the data strongly contradict the prior information.
We have chosen $d = 6$ because  
extensive numerical experimentation shows this is sufficiently large. 
Let ${\cal D}$ denote the class of pairs of functions $(f_o, f_e)$ that satisfy these requirements.

Let ${\cal C}$ denote the subset of ${\cal D}$ such that $(f_o, f_e)$ 
is fully specified by the vectors $\big(f_o(1), f_o(2), \dots, f_o(5) \big)$ and
$\big(f_e(0), f_e(1), \dots, f_e(5) \big)$ as follows. 
By assumption, 
$\big(f_o(-1), f_o(-2), \dots, f_o(-5) \big)
= \big(-f_o(1), -f_o(2), \dots, -f_o(5) \big)$
and 
$\big( f_e(-1), \dots, f_e(-5) \big) \newline
= \big( f_e(1), \dots, f_e(5) \big)$.
The values of $f_o(x)$ and $f_e(x)$ for any $x \in [-6, 6]$ are found by 
natural 
cubic spline interpolation for the given values of $f_o(j)$ and $f_e(j)$ for 
$j = -6, -5, \dots, 0, 1, \dots, 5, 6$. We will numerically compute 
$(f_o, f_e) \in {\cal C}$ 
such that $J\big(f_o, f_e; \sigma_{\varepsilon}, \delta\big)$ has 
minimum coverage probability $1 - \alpha$ and scaled expected length
that (a) is substantially less than 1 when the prior information is correct and (b) has maximum (over the parameter space) that is not too much larger than 1.

We now describe the numerical constrained optimization method used to find the pair of functions $(f_o, f_e) \in {\cal C}$ satisfying these conditions.
Let $\gamma = \xi \, N^{1/2} / \sigma_{\varepsilon}$, a scaled version of the non-exogeneity parameter $\xi$.
By Theorem B.1, stated in Appendix B, for any given $(f_o, f_e) \in {\cal D}$,
the conditional coverage probability 
$P \big(b \in J(f_o, f_e; \sigma_{\varepsilon}, \delta) \, \big| \, x \big)$
is a function of $(\gamma, \delta$). We therefore denote this coverage 
probability by $CP(\gamma, \delta; f_o, f_e)$.
We define the conditional scaled expected length of 
$J(f_o, f_e; \sigma_{\varepsilon}, \delta)$
as 
\begin{equation*}
\frac{E\big (\text{length of} \ J(f_o, f_e; \sigma_{\varepsilon}, \delta)
\, \big| \, x	\big)}
{\text{length of} \ L(\sigma_{\varepsilon}, 1-\alpha) }.
\end{equation*}
By Theorem B.1, stated in Appendix B, for any given $(f_o, f_e) \in {\cal D}$,
this scaled expected length 
is a function of $(\gamma, \delta$). We therefore denote this 
scaled expected length by $SEL(\gamma, \delta; f_o, f_e)$.

Suppose for the moment that $\varphi$, satisfying $0 \le \varphi \le 1$,
is given.
Numerically compute the pair of functions $(f_o, f_e) \in {\cal C}$ that
minimizes the objective function
\begin{equation}
\label{CriterionVarphi}
(1 - \varphi) \big(SEL(\gamma = 0, \delta; f_o, f_e) - 1\big)
+ \varphi \int_{-\infty}^{\infty} \big(SEL(\gamma, \delta; f_o, f_e) - 1\big)
\, d\gamma,
\end{equation}
subject to the inequality constraint
that 
$CP(\gamma, \delta; f_o, f_e) \ge 1 - \alpha$ for all $\gamma \in \mathbb{R}$.
This numerical constrained optimization is made feasible by the computationally 
convenient expressions for $CP(\gamma, \delta; f_o, f_e)$, which is an even function of $\gamma$, and 
the objective function \eqref{CriterionVarphi} given in Appendix A.
For the computations, the coverage inequality constraint $CP(\gamma, \delta; f_o, f_e) \geq 1 - \alpha$ for all $\gamma \geq 0$ is replaced by $CP(\gamma, \delta; f_o, f_e) \geq 1 - \alpha$ for a well chosen finite set of nonnegative values of $\gamma$. 
Let $\big(f_o^*(\, \cdot \,;\delta, \varphi), f_e^*(\, \cdot \,;\delta,\varphi) \big)$ denote the value of $(f_o, f_e) \in {\cal C}$ that results from this numerical computation.

For $\varphi = 1$ this numerical computation recovers the confidence
interval $L(\sigma_{\varepsilon}, 1 - \alpha)$ for $b$, based on the {\sl fixed effects model} and with coverage probability $1 - \alpha$. As $\varphi$ decreases from
this value, this computation puts increasing weight on achieving a small value
of $SEL(\gamma = 0, \delta; f_o, f_e)$, which results in a smaller value of 
$SEL\big(\gamma = 0, \delta; f_o^*(\, \cdot \,;\delta, \varphi),f_e^*(\, \cdot \,;\delta, \varphi) \big)$ i.e. an improved confidence interval performance
when the prior information that $\gamma = 0$ is correct. However, as $\varphi$ decreases
\begin{equation*}
\max_{\gamma} SEL\big(\gamma, \delta; f_o^*(\, \cdot \,;\delta, \varphi),f_e^*(\, \cdot \,;\delta, \varphi)  \big)
\end{equation*}
increases i.e. the performance of the confidence interval when the prior information happens to be incorrect is degraded. Numerical experimentation revealed that a reasonable tradeoff between this improvement and degradation in performance resulted when $\varphi$ was chosen such that the ``gain'' when the prior information is correct,
as measured by 
\begin{equation}
\label{SEL_gain}
1 - \bigg(SEL\big(\gamma = 0, \delta; f_o^*(\, \cdot \,;\delta, \varphi),f_e^*(\, \cdot \,;\delta, \varphi) \big) \bigg)^2,
\end{equation}
is equal to the maximum possible ``loss'' when the prior information happens to be incorrect, as measured by 
\begin{equation}
\label{SEL_loss}
\left(\max_{\gamma} SEL \big(\gamma, \delta;f_o^*(\, \cdot \,;\delta, \varphi),f_e^*(\, \cdot \,;\delta, \varphi) \big) \right)^2 - 1.
\end{equation}
Clearly, the value of $\varphi$ chosen in this way is determined by $\delta$ (assuming that $x$ is given) and so we denote it by 
$\varphi^*(\delta)$.
In other words, the confidence interval constructed assuming that 
$(\sigma_{\varepsilon}, \delta)$ is known is
\begin{equation*}
CI(\sigma_{\varepsilon}, \delta) = J\Big(f_o^*(\, \cdot \,;\delta, \varphi^*(\delta)), \, f_e^*(\, \cdot \,;\delta, \varphi^*(\delta)); \, \sigma_{\varepsilon}, \, \delta \Big).
\end{equation*} 

\begin{figure}[h]
\centering
\includegraphics[scale = 0.7]{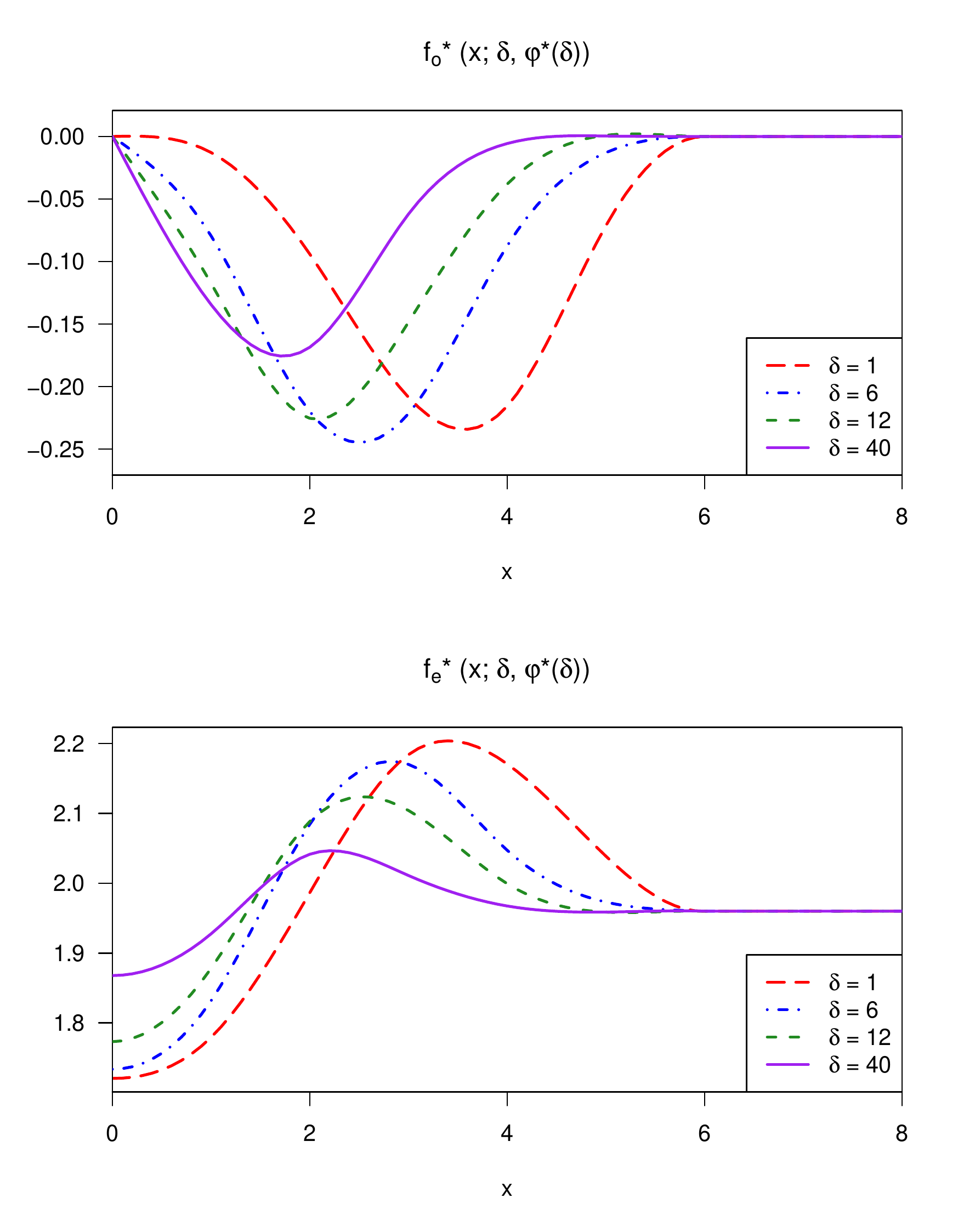}
\vspace{-0.3cm}
\caption{Graphs of the functions $f_o^*(\, \cdot \,;\delta, \varphi^*(\delta))$ (top panel) and $f_e^*(\, \cdot \,;\delta, \varphi^*(\delta))$ (bottom panel) for 
	$1 - \alpha = 0.95$ and each $\delta \in \{1, 6, 12, 40\}$, for the airfare data.}
\label{f*_airfare}
\end{figure}

For $\delta \in \{1, 6, 12, 40\}$, Figure \ref{f*_airfare} presents graphs of the functions
$f_o^*(\, \cdot \,;\delta, \varphi^*(\delta))$ 
(top panel) and 
$f_e^*(\, \cdot \,;\delta, \varphi^*(\delta))$ 
(bottom panel) for $1 - \alpha = 0.95$, for the airfare data.
Because $f_o^*(\, \cdot \,;\delta, \varphi^*(\delta))$ is an odd function and 
$f_e^*(\, \cdot \,;\delta, \varphi^*(\delta))$ 
is an even function, we present these graphs only for $x \geq 0$.

\FloatBarrier

\begin{figure}[h]
	\centering
	\includegraphics[scale = 0.8]{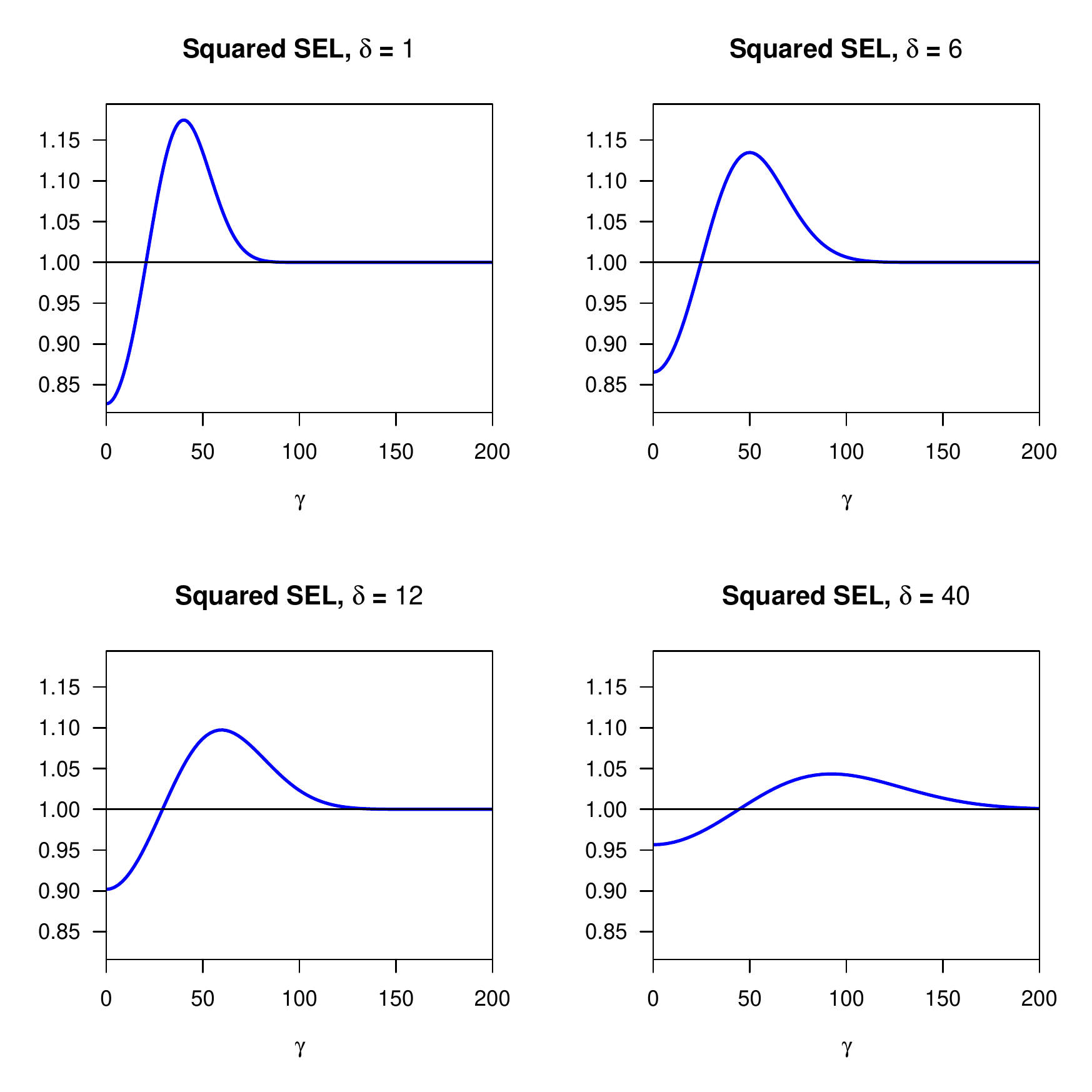}
	\vspace{-0.3cm}
	\caption{Graphs of the squared scaled expected length of $CI(\sigma_{\varepsilon}, \delta)$ as a function of $\gamma$ for $\delta \in \{1, 6, 12, 40\}$ and $1 - \alpha = 0.95$.}
	\label{SquaredSEL_f*_airfare}
\end{figure}

For the airfare data and $1 - \alpha = 0.95$, the infimum and supremum of the coverage probability
of $CI(\sigma_{\varepsilon}, \delta)$,
conditional on $x$, are  0.9500 and 0.9507, respectively. In other words, the coverage probability of this interval, conditional on $x$, is very close to $1 - \alpha = 0.95$
for all $(\gamma, \delta)$.
For the airfare data, Figure \ref{SquaredSEL_f*_airfare} presents graphs of the squared scaled expected length of $CI(\sigma_{\varepsilon}, \delta)$ as a function of $\gamma$ for each $\delta \in \{1, 6, 12, 40\}$ and $1 - \alpha = 0.95$.  We can see from this figure that, for each of these values of $\delta$, the minimum squared scaled expected length is below 1 at $\gamma = 0$ (i.e. when the uncertain prior information is correct), the maximum squared scaled expected length is not too much greater than 1, and the squared scaled expected length approaches 1 as $\gamma$ increases.  This last property is a consequence of the fact that, by construction, $CI(\sigma_{\varepsilon}, \delta)$ reverts to the confidence interval $L(\sigma_{\varepsilon}, 1 - \alpha)$, based on the \textsl{fixed effects model}, when the data strongly contradict the prior information.

\FloatBarrier

\subsection{Second step: replace the true parameter value 
	$\boldsymbol{(\sigma_{\varepsilon}, \delta)}$ by an \newline estimator $\boldsymbol{(\widehat{\sigma}_{\varepsilon}, \widehat{\delta})}$}
\label{sigma_delta_unknown}

We use the same estimator $(\widehat{\sigma}_{\varepsilon}, \widehat{\delta})$ of 	$(\sigma_{\varepsilon}, \delta)$ as Kabaila \textit{et al} (2017). The  motivation for this estimator is provided in Section 2.1 of  Kabaila \textit{et al} (2017).  
Define $\overline{y}_i = T^{-1} \sum_{t=1}^T y_{it}$ and $\overline{\varepsilon}_i = T^{-1} \sum_{t=1}^T \varepsilon_{it}$.  
Let 
$r_{it} = (y_{it} - \overline{y}_i) - \widehat{b}_W (x_{it} - \overline{x}_i)$.
The estimator of $\sigma_{\varepsilon}^2$ that we use is 
\begin{equation*}
\widehat{\sigma}_{\varepsilon}^2 = \dfrac{1}{N(T-1)} \sum_{i=1}^N \sum_{t=1}^T r_{it}^2.
\end{equation*}
Let $\widehat{a}$ denote the GLS estimator of $a$ based on the model \eqref{CRE_model2_l2}.  
Define $\widetilde{r}_i = \overline{y}_i - (\widehat{a} + \widehat{b}_B \, \overline{x}_i)$.  
The estimator of $\delta$ that we use is $\widehat{\delta} = \widehat{\sigma}_{\eta}^2 \, / \, \widehat{\sigma}_{\varepsilon}^2$, where
\begin{equation*}
\widehat{\sigma}_{\eta}^2 = \dfrac{1}{N} \sum_{i=1}^N \widetilde{r}_i^2 - \dfrac{\widehat{\sigma}_{\varepsilon}^2}{T}.
\end{equation*}

The second step in the construction of the new confidence interval is to replace $(\sigma_{\varepsilon}, \delta)$ 
by the estimator $(\widehat{\sigma}_{\varepsilon}, \widehat{\delta})$
in $CI(\sigma_{\varepsilon}, \delta)$.  
In other words, we use the plug-in principle (Efron, 1998, Section 5).
The resulting confidence interval is 
$CI(\widehat{\sigma}_{\varepsilon}, \widehat{\delta})$.  
As proved in Appendix B, the conditional coverage probability and the scaled expected length of this confidence interval are both functions of $(\gamma, \delta)$.

As noted in Subsection 2.1, for $(\sigma_{\varepsilon}, \delta)$ assumed known,
the infimum over $(\gamma, \delta)$ of the coverage probability of 
$CI(\sigma_{\varepsilon}, \delta)$, conditional on $x$, is 0.9500 (accurate to four digits after the decimal point) which is
very close to $1 - \alpha = 0.95$. As evidenced by Figure 1, this very desirable property is, to a large extent, inherited by the new confidence interval obtained through the application of the plug-in principle. A comparison of Figures \ref{fig:SEL_airfare} and \ref{SquaredSEL_f*_airfare} shows that the squared scaled expected length of 
$CI(\widehat{\sigma}_{\varepsilon}, \widehat{\delta})$
is very close to the squared scaled expected length of $CI(\sigma_{\varepsilon}, \delta)$.  To summarize, for the airfare data the plug-in principle works well.




\section{SIMULATION METHODS USED TO EVALUATE THE CONDITIONAL COVERAGE AND SCALED EXPECTED LENGTH OF THE NEW CONFIDENCE INTERVAL}
\label{sim_method}

In this section we briefly describe the two main ideas employed in the simulation methods that are used to evaluate the conditional coverage and the scaled expected length of the new confidence interval.  A detailed description of these simulation methods is given in Appendix C.

The first main idea is as follows.  For each given value of $\delta$, the functions $f_o^*(\, \cdot \,;\delta, \varphi^*(\delta))$ and $f_e^*(\, \cdot \,;\delta, \varphi^*(\delta))$ are found using the method described in Subsection \ref{subsec_CI_sig_del_known}, with computation time of the order of 1 minute. To make the simulations computationally feasible, we precompute the 
functions $f_o^*\big(\, \cdot \,;\delta, \varphi^*(\delta)\big)$ and $f_e^*\big(\, \cdot \,;\delta, \varphi^*(\delta) \big)$ for a grid of 11 values of 
$\delta$, denoted by $\delta_1, \dots, \delta_{11}$. For any given $\delta \in [\delta_j, \delta_{j+1}]$ and given $x \in [-6, 6]$, the value of 
$f_o^*\big(x;\delta, \varphi^*(\delta)\big)$ is obtained by linear interpolation in $\delta$
from $f_o^*\big(x;\delta_j, \varphi^*(\delta_j) \big)$ and
$f_o^*\big(x;\delta_{j+1}, \varphi^*(\delta_{j+1}) \big)$.

The grid of 11 values of $\delta$ is obtained as follows.
Let
\begin{equation}
\label{eqn_rho}
\rho(\delta) = - \left( \dfrac{r(x)}{r(x) + \delta + T^{-1}} \right)^{1/2},
\end{equation}
where $r(x) = \SSB / \SSW$.
It can be shown that $\rho(\delta)$ is the correlation between $\widehat{b}_W$ and $\widehat{b}_B - \widehat{b}_W$.  Because $r(x)$ and $\delta$ are nonnegative,  $\rho(\delta) \in (-1, 0]$. To obtain the grid of values of $\delta$ we begin
with the grid $\rho_1 = 0, \rho_2 = -0.1, \rho_3 = -0.2, \dots, \rho_{10} = -0.9,
\rho_{11} = -0.97$ of values of $\rho$. We then find $\delta_j$ by solving 
for $\delta$ in $\rho(\delta) = \rho_j$ ($j=1, \dots, 11$).
Our grid of values of $\delta$ is, then, $\delta_1, \dots, \delta_{11}$.
Note that when $\rho(\delta) = 0$, $CI(\sigma_{\varepsilon}, \delta)$  is equal to 
$L(\sigma_{\varepsilon}, 1 - \alpha)$.

The second main idea is as follows.   Define $\eta_i^{\dag} = \eta_i / \sigma_{\eta}$ and $\varepsilon_{it}^{\dag} = \varepsilon_{it}/\sigma_{\varepsilon}$
for $i = 1, \dots, N$ and $t=1, \dots, T$.  Observe that the $\eta_i^{\dag}$'s and $\varepsilon_{it}^{\dag}$'s are i.i.d. $N(0, 1)$.  It follows from Kabaila \textit{et al} (2017, Appendix A.3) that the conditional coverage probability and scaled expected length of the new confidence interval $CI(\widehat{\sigma}_{\varepsilon}, \widehat{\delta})$ can be expressed in terms of the $\eta_i^{\dag}$'s, $\varepsilon_{it}^{\dag}$'s, $\delta$, $\gamma$ and $x$.  We use the $\eta_i^{\dag}$'s and $\varepsilon_{it}^{\dag}$'s to drive the simulations, thereby removing the need to specify values for either $\sigma_{\varepsilon}$ or $\sigma_{\eta}$.

\section{ANALYSIS OF THE ROBUSTNESS OF THE NEW CONFIDENCE INTERVAL}

The new confidence interval is constructed assuming
 that the $\eta_i$'s and the $\varepsilon_{it}$'s are independent, with the $\eta_i$'s i.i.d $N(0, \sigma^2_{\eta})$ and the $\varepsilon_{it}$'s are i.i.d. $N(0, \sigma^2_{\varepsilon})$. 
As previously, let $\varepsilon_{it}^{\dag} = \varepsilon_{it}/\sigma_{\varepsilon}$ and $\eta_i^{\dag} = \eta_i / \sigma_{\eta}$, so that the $\varepsilon_{it}^{\dag}$'s and the $\eta_i^{\dag}$'s are i.i.d. 
 $N(0, 1)$. 
  It follows from Kabaila, Mainzer and Farchione (2017, Appendix A.3) that the conditional coverage probability and scaled expected length of the new confidence interval $CI(\widehat{\sigma}_{\varepsilon}, \widehat{\delta})$ can be expressed in terms of the $\eta_i^{\dag}$'s, $\varepsilon_{it}^{\dag}$'s, $\delta$, $\gamma$ and $x$.

  In the Supplementary Material we suppose that the $\eta_i^{\dag}$'s and $\varepsilon_{it}^{\dag}$'s are still independent, but with the 
  $\eta_i^{\dag}$'s identically distributed with a skewed standardized Student's $t$-distribution and the $\varepsilon_{it}^{\dag}$'s
  identically distributed with a, possibly different, skewed standardized Student's $t$-distribution.
  For the airfare data,
  we assess the impact of this change of distributions on the following three properties of the new confidence interval: the minimum conditional coverage
  probability, the scaled expected length for $\gamma = 0$ and the scaled expected length, maximized over $\gamma$.  As noted earlier, this confidence interval is constructed assuming that the $\varepsilon_{it}^{\dag}$'s and  $\eta_i^{\dag}$'s are i.i.d. $N(0, 1)$. In other words, we assess the robustness of
  these properties of the new confidence interval 
  to replacement of the distributions of the $\eta_i^{\dag}$'s and $\varepsilon_{it}^{\dag}$'s by skewed standardized Student's $t$-distributions. 
  As shown in the Supplementary Material, these properties are very 
  robust to this replacement.

 \section{DISCUSSION}
 
 Until the present paper, the applied econometrician has been faced with 
 choosing between the following two options. The first of these options is to
 carry out a Hausman pretest for exogeneity followed by the construction
 of the confidence interval for the slope parameter, leading to a confidence interval
 with unacceptably poor coverage properties. The second of these options is to always construct the
 confidence interval using the {\sl fixed effects model} i.e. to always avoid
 using the Hausman pretest.  
Neither of these options would seem to be particularly attractive
 to the applied econometrician.

 The new confidence interval described in the present paper provides an intermediate between these two options
 and, in a sense, combines the best features of both.
 To an excellent approximation, the new confidence interval 
 has the desired minimum coverage probability. 
 The test statistic for the Hausman pretest enters into the expression for 
 the new confidence interval. 
 Furthermore, this confidence interval reduces 
 to the confidence interval obtained using the {\sl fixed effects model}
 when the value of the test statistic for the Hausman pretest strongly contradicts the 
 prior information.

 We have defined the ``gain'' of the new confidence interval to be 1 minus its squared scaled expected length, when the prior information is correct (i.e. when the covariate is exogenous). We have also defined the maximum possible ``loss'' to be the maximum of the squared scaled expected length minus 1, when the prior information happens to be incorrect. For the new confidence interval, we have set the ``gain''  equal to the maximum possible ``loss''. However, if the data strongly contradict the prior information then the ``loss'' is negligible. 
 
 Put another way, the new confidence interval does not dominate the confidence interval, with the same confidence coefficient, obtained using the \textsl{fixed effects model}. Indeed, the fact that the application of the plug-in principle works so well and the admissibility result of Kabaila, Giri and Leeb (2010, Section 4) suggest that it is impossible to find a confidence interval that dominates the confidence interval obtained using the \textsl{fixed effects model}.
 Overall, the new confidence interval should only be used when there is some reasonable confidence (although not certainty) in the prior information that the covariate is exogenous. Of course, the decision as to whether or not the new confidence interval will be used must be made prior to examination of the vector of observed responses.
 
 

The new confidence interval is constructed using the method of Kabaila and Giri (2009), as described in Appendix A. The computation of this confidence interval  
 has previously been carried out using the numerical constrained 
 optimization function {\tt fmincon} in the {\tt MATLAB Optimization Toolbox}. 
 The {\tt MATLAB programs} for the computation of this confidence interval and its conditional coverage and scaled expected length have been translated, with improvements, to {\tt R} programs, which will be made available in an {\tt R} package.
 
\section*{Acknowledgement}

This work was supported by an Australian Government Research Training Program Scholarship.

\bigskip

\noindent {\large \textsl{REFERENCES}}

\medskip

\rf Abeysekera, W. and P. Kabaila (2017) Optimized recentered confidence spheres for the multivariate normal mean.  \textsl{Electronic Journal of Statistics} 11, 1798--1826.

\smallskip 

\rf Bickel, P.J. (1984) Parametric robustness: small biases can be
worthwhile. \textsl{Annals of Statistics} 12, 864--879.

\smallskip

\rf Bickel, P.J. and K.A. Doksum (1977) \textsl{Mathematical Statistics, Basic Ideas and Selected Topics} Holden-Day.

\smallskip
 
\rf Casella, G. and J.T. Hwang (1983) Empirical Bayes confidence sets for the mean of a multivariate normal distribution. \textsl{Journal of the American Statistical Association} 78, 688--698.

\smallskip

\rf Casella, G. and J.T. Hwang (1987) Employing vague prior information in the construction of confidence sets. \textsl{Journal of Multivariate Analysis} 21, 79--104.

\smallskip

\rf Choe, J. (2008) Income inequality and crime in the United States.  \textsl{Economics Letters} 101, 31--33.

\smallskip

\rf Cohen, A. (1972) Improved confidence intervals for the variance of a normal distribution.
\textsl{Journal of the American Statistical Association} 67, 382--387.

\smallskip

\rf Efron, B. (1998) R.A. Fisher in the 21st century. \textsl{Statistical Science}
13, 95--112.

\smallskip

\rf Efron, B. (2006) Minimum volume confidence regions for a multivariate normal mean vector. \textsl{Journal of the Royal Statistical Society, Series B} 68, 655--670.

\smallskip

\rf Giri, K. (2008) \textsl{Confidence Intervals in Regression Utilizing Prior Information.} Unpublished PhD thesis.

\smallskip

\rf Goutis, C. and G. Casella (1991) Improved invariant confidence intervals for a normal
variance. \textsl{Annals of Statistics} 19, 2015--2031.

\smallskip

\rf Guggenberger, P. (2010) The impact of a Hausman pretest on the size of a hypothesis test: the panel data case. \textsl{Journal of Econometrics} 156, 337--343.

\smallskip

\rf Hastings, J.S. (2004) Vertical relationships and competition in retail gasoline markets: Empirical evidence from contract changes in Southern
California. \textsl{American Economic Review} 94, 317--328.

\smallskip

\rf Hausman, J. A. (1978) Specification tests in econometrics. \textsl{Econometrica} 46, 1251--1271.

\smallskip

\rf Hodges, J.L. and E.L. Lehmann  (1952) The use of previous
experience in reaching statistical decisions. \textsl{Annals of
	Mathematical Statistics} 23, 396--407.

\smallskip

\rf Jackowicz, K., O. Kowalewski and \L{}. Koz\l{}owski (2013) The influence of political factors on commercial banks in Central European countries.  \textsl{Journal of Financial Stability} 9, 759--777.

\smallskip

\rf Kabaila, P. (2009) The coverage properties of confidence regions after model selection. \textsl{International Statistical Review} 77, 405--414.

\smallskip

\rf Kabaila, P. and K. Giri (2009) Confidence intervals in regression utilizing prior information. \textsl{Journal of Statistical Planning and Inference} 139, 3419--3429.

\smallskip

\rf Kabaila, P., K. Giri and H. Leeb (2010) Admissibility of the usual confidence interval in linear regression.  \textsl{Electronic Journal of Statistics} 4, 300--312.

\smallskip

\rf Kabaila, P. and H. Leeb (2006) On the large-sample minimal coverage probability of confidence intervals after model selection.  \textsl{Journal of the American Statistical Association} 101, 619--629.

\smallskip

\rf Kabaila, P., R. Mainzer and D. Farchione (2015) The impact of a Hausman pretest, applied to panel data, on the coverage probability of confidence intervals. \textsl{Economics Letters} 131, 12--15.

\smallskip

\rf Kabaila, P., R. Mainzer and D. Farchione (2017) Conditional assessment of the impact of a Hausman pretest on confidence intervals. \textsl{Statistica Neerlandica} 

\smallskip

\rf Kempthorne, P.J. (1983) Minimax-Bayes compromise estimators. In {\it 1983 Business and 
	Economic Statistics Proceedings of the American Statistical Association}, pp.568--573,
Washington DC.

\smallskip

\rf Kempthorne, P.J. (1987) Numerical specification of discrete least favourable
distributions. \textsl{SIAM Journal of Scientific and Statistical Computing} 8,
171--184.

\smallskip

\rf Kempthorne, P.J. (1988) Controlling risks under different loss functions:
the compromise decision problem. 
\textsl{Annals of Statistics} 16, 1594--1608.

\smallskip

\rf Koopmans, T.C. (1937) \textsl{Linear Regression Analysis of Economic Time Series}. Haarlem, Netherlands: Bohn.

\smallskip

\rf Leamer, E. E. (1978) \textsl{Specification Searches: Ad Hoc Inference with Nonexperimental Data.}  New York: Wiley.

\smallskip

\rf Leeb, H. and B. P\"otscher (2005) Model selection and inference: facts and fiction.  \textsl{Econometric Theory} 21, 21--59.

\smallskip

\rf Papatheodorou, A. and Z. Lei (2006) Leisure travel in Europe and airline business models: A study of regional airports in Great Britain.  \textsl{Journal of Air Transport Management} 12, 47--52.

\smallskip

\rf Pratt, J.W. (1961) Length of confidence intervals. \textsl{Journal of the American Statistical Association} 56, 549--657.

\smallskip

\rf Smith, N., V. Smith and M. Verner (2006) Do women in top management affect firm performance? A panel study of 2,500 Danish firms.  \textsl{International Journal of Productivity and Performance Management} 55, 569--593.

\smallskip

\rf Stanca, L. (2006) The effects of attendance on academic performance: panel data evidence for Introductory Microeconomics.  \textsl{The Journal of Economic Education} 37, 251--266.

\smallskip

\rf Stein, C. (1962) Confidence sets for the mean of a multivariate normal
distribution.
\textsl{Journal of the Royal Statistical Society, Series B} 9, 1135--1151.

\smallskip

\rf Tseng, Y.L. and L.D. Brown (1997) Good exact confidence sets for a multivariate normal mean.
\textsl{Annals of Statistics} 5, 2228--2258.

\smallskip

\rf Wooldridge, J. M. (2013), \textsl{Introductory Econometrics: a Modern Approach,} 5th edtion.  Ohio: South-Western.

\bigskip

\begin{center}
\bf{\large{APPENDIX A}}
\end{center}

\renewcommand{\theequation}{A.\arabic{equation}}

\noindent \textbf{Standard transformation of the model \eqref{CRE_model2_l2} when $\boldsymbol{(\sigma_{\varepsilon}, \delta)}$ is assumed to be known}

\medskip

\noindent Express the model \eqref{CRE_model2_l2} in matrix form as 
\begin{equation}
\label{matrix_model}
Y = X \beta + u,
\end{equation}
where 
\begin{equation*}
Y = \begin{bmatrix} y_{11} \\ \vdots \\ y_{1T} \\ \vdots \\ y_{N1} \\ \vdots \\ y_{NT} \end{bmatrix},
\, \, 
X = \begin{bmatrix} 1 & x_{11} - \overline{x}_1 & \overline{x}_1 \\  \vdots \\ 1 & x_{1T} - \overline{x}_1 & \overline{x}_1 \\ \vdots \\ 1 & x_{N1} - \overline{x}_N & \overline{x}_N \\ \vdots \\ 1 & x_{NT} - \overline{x}_N & \overline{x}_N \end{bmatrix},
\, \, 
\beta = \begin{bmatrix} a \\ b_W \\ b_B \end{bmatrix}
\, \, 
\text{ and } u = \begin{bmatrix} \eta_1 + \varepsilon_{11} \\ \vdots \\ \eta_1 + \varepsilon_{1T} \\ \vdots \\ \eta_N + \varepsilon_{N1} \\ \vdots \\ \eta_N + \varepsilon_{NT} \end{bmatrix}.
\end{equation*}
Remember, from Subsection \ref{subsec_CI_sig_del_known}, that $b_W = b$ and $b_B = b + \xi$.

The covariance matrix of $u$ is $\sigma^2_{\varepsilon} \, C(\delta)$, where $C(\delta)$ is an $NT \times NT$ block diagonal matrix with $N$ identical block diagonal elements each of which is the $T \times T$ matrix $I_T + \delta \, e_T \, e_T^{\top}$, where $e_T$ is a $T$-vector of 1's. 
Also, $\big(C(\delta)\big)^{-1}$ is the $NT \times NT$ block diagonal matrix with $N$ identical block diagonal elements, each of which is the $T \times T$ matrix $I_T - (\delta / (1 + \delta T)) \, e_T \, e_T^{\top}$. 
Premultiplying \eqref{matrix_model} by $\big(C(\delta)\big)^{-1/2}$ gives
\begin{equation}
\label{transformed_matrix_model}
\widetilde{Y} = \widetilde{X} \beta + \widetilde{u},
\end{equation}
where $\widetilde{Y} = \big(C(\delta)\big)^{-1/2} \,Y$, $\widetilde{X} = \big(C(\delta)\big)^{-1/2} \, X$ and $\widetilde{u} = \big(C(\delta)\big)^{-1/2} \, u$.  Then $\text{Cov}(\widetilde{u}) = \sigma^2_{\varepsilon} \, I_{NT}$.  
Now assume that $(\sigma_{\varepsilon}, \delta)$ is known, so that  \eqref{transformed_matrix_model}
describes a linear regression model with known random error variance $\sigma^2_{\varepsilon}$.
The parameter of interest is $b_W = b$ and we suppose that we have uncertain prior information that $\xi = b_B - b_W = 0$.  

\medskip


\medskip

\noindent \textbf{Description of the Kabaila and Giri confidence interval for known error variance}

\medskip

\noindent Suppose that 
\begin{equation}
\label{ModelUsedByKG}
\widetilde{Y} = \widetilde{X} \beta + \widetilde{u},
\end{equation}
where $\widetilde{Y}$ is a random $n$-vector of responses, $\widetilde{X}$ is a known
$n \times p$ matrix ($n > p$) with linearly independent columns, $\beta$ is an 
unknown parameter $p$-vector and $\widetilde{u} \sim N(0, \sigma^2 I_n)$
where $\sigma^2$ is known. Also suppose that the parameter of 
interest is $\theta = \widetilde{a}^{\top} \beta$, where $\widetilde{a}$ is a specified non-zero
$p$-vector. Let $\xi = \widetilde{c}^{\top} \beta - \widetilde{t}$, where $\widetilde{c}$ and $\widetilde{t}$ are specified and  $\widetilde{a}$ and 
$\widetilde{c}$ are linearly independent. Suppose that we have uncertain
prior information that $\xi = 0$. We will construct a confidence interval for
$\theta$ that has minimum coverage probability $1 - \alpha$ and utilizes this
uncertain prior information through the expected length properties
described later in this subsection.

Let $\widehat{\beta}$ denote the OLS estimator of $\beta$, based on the model
\eqref{ModelUsedByKG}. Also let  
$\widehat{\theta} = \widetilde{a}^{\top} \widehat{\beta}$
and $\widehat{\xi} = \widetilde{c}^{\top} \widehat{\beta} - \widetilde{t}$.
Define $v_{\theta} = \text{var}(\widehat{\theta}) / \sigma^2$, 
$v_{\xi} = \text{var}(\widehat{\xi}) / \sigma^2$ and
$\rho = \text{corr}(\widehat{\theta}, \widehat{\xi})$. Note that 
$v_{\theta}$, $v_{\xi}$ and $\rho$ are known. Let
$\psi = \xi / (\sigma v_{\xi}^{1/2})$ and 
$\widehat{\psi} = \widehat{\xi} / (\sigma v_{\xi}^{1/2})$.
The scenario described in the last paragraph of the first subsection of this appendix 
arises from the particular case that $p = 3$, 
$\widetilde{a} = (0, 1, 0)$, $\widetilde{c} = (0, -1, 1)$ and $(\sigma_{\varepsilon}, \delta)$ are known.
For that scenario, $\sigma^2 = \sigma^2_{\varepsilon}$,
$v_{\theta} = 1 / \SSW$, 
\begin{equation*}
 v_{\xi} =\frac{1}{\SSW} \left(\dfrac{r(x) + \delta + T^{-1}}{r(x)} \right),
\end{equation*}
and
\begin{equation*}
\rho = - \left( \dfrac{r(x)}{r(x) + \delta + T^{-1}} \right)^{1/2}.
\end{equation*}
We define $\gamma = \xi N^{1/2} / \sigma_{\varepsilon}$. It follows that 
\begin{equation*}
\psi = \frac{\gamma}{ \left(\dfrac{N}{\SSW} \left(\dfrac{r(x) + \delta + T^{-1}}{r(x)} \right )
	\right)^{1/2}},
\end{equation*}
and $\widehat{\psi} = h(\sigma_{\varepsilon}, \delta)$.

We consider confidence intervals of the form
\begin{equation*}
J(f_o, f_e) 
= \left[\widehat{\theta} - v_{\theta}^{1/2} \sigma f_o \big(\widehat{\psi} \, \big)
\pm v_{\theta}^{1/2} \sigma f_e \big(\widehat{\psi} \, \big)
\right],
\end{equation*}
where $f_o: \mathbb{R} \rightarrow \mathbb{R}$ is an odd continuous function
and $f_e: \mathbb{R} \rightarrow [0, \infty)$ is an even continuous function.
In addition, we require that $f_o(x) = 0$ and $f_e(x) = z_{1-\alpha/2}$ for all $|x| \ge d$, where 
$d$ is a (sufficiently large) specified positive number.
We will numerically compute the functions $f_o^*$ and $f_e^*$ such that 
$J(f_o^*, f_e^*)$ has 
minimum coverage probability $1 - \alpha$ and the 
desired expected length properties.

Kabaila and Giri (2009) deal with the more difficult case that $\sigma^2$ is
unknown.  They provide computationally convenient formulae for the  coverage probability 
and scaled expected 
length of confidence intervals of similar form to $J(f_o, f_e)$ when $\sigma^2$ is 
unknown. They also provide a computationally convenient formula for the 
objective function used in the numerical constrained optimization (described below) to find the functions $f_o^*$ and $f_e^*$.

These formulae simplify in the present case that $\sigma^2$ is 
known. For the sake of brevity, we omit the derivations of the following results.
The coverage probability 
$P \big(\theta \in J(f_o, f_e) \big)$
is an even function of $\psi$ which we denote by $CP(\psi; f_o, f_e)$. A computationally convenient 
formula for this coverage probability, given by Giri (2008, Subsection 4.3), is
\begin{equation*}
CP(\psi; f_o, f_e) = 1 - \alpha + \int_{-d}^d \left(k(w) - k^{\dagger}(w) \right) \phi(w - \psi) \, dw,
\end{equation*}
where
\begin{equation*}
k(w) = \Phi \left( \dfrac{f_o(w) + f_e(w) - \rho (w - \psi)}{\sqrt{1 - \rho^2}} \right) - \Phi \left( \dfrac{f_o(w) - f_e(w) - \rho(w - \psi)}{\sqrt{1 - \rho^2}}\right)
\end{equation*}
and
\begin{equation*}
k^{\dag}(w) = \Phi \left( \dfrac{z_{1-\alpha/2} - \rho (w - \psi)}{\sqrt{1 - \rho^2}} \right) - \Phi \left( \dfrac{-z_{1-\alpha/2} - \rho (w - \psi)}{\sqrt{1 - \rho^2}} \right).
\end{equation*}

The standard $1 - \alpha$ confidence interval for $\theta$ is 
\begin{equation*}
I = \left[\widehat{\theta} - v_{\theta}^{1/2} \sigma z_{1-\alpha/2}, \,
\widehat{\theta} + v_{\theta}^{1/2} \sigma z_{1-\alpha/2}
\right].
\end{equation*}
We define the scaled expected length of $J(f_o, f_e)$ to be 
\begin{equation*}
\frac{E\big(\text{length of} \ J(f_o, f_e)\big)}
{\text{length of} \ I}.
\end{equation*}
This scaled expected length 
is an even function of $\psi$ which we denote  by $SEL(\psi; f_o, f_e)$.
A computationally convenient formula for this scaled expected length, given by Giri (2008, Subsection 4.3), is 
\begin{equation*}
SEL(\psi; f_o, f_e) = 1 + \frac{1}{z_{1-\alpha/2}} \int_{-d}^d \left( f_e(w) - z_{1-\alpha/2} \right) \phi(w - \psi) \, dw.
\end{equation*}

A computationally 
convenient formula for the objective function
\begin{equation*}
(1 - \varphi) \big(SEL(\psi = 0; f_o, f_e) - 1 \big)
+ \varphi \int_{-\infty}^{\infty} \big(SEL(\psi; f_o, f_e) - 1 \big)
\, d\psi,
\end{equation*}
is
\begin{equation*}
\frac{2}{z_{1 - \alpha/2}} 
\int_0^d \big(f_e(w) - z_{1 - \alpha/2}\big) \big((1 - \varphi)\phi(w) + \varphi \big) \, dw.
\end{equation*}
We numerically compute the pair of functions $(f_o, f_e) \in {\cal C}$ that
minimizes this objective function
subject to the inequality constraint
that 
$CP(\psi; f_o, f_e) \ge 1 - \alpha$ for all $\psi \ge 0$.
For these computations, this inequality constraint  is replaced by 
$CP(\psi; f_o, f_e) \geq 1 - \alpha$ for a well chosen finite set of nonnegative values of $\psi$. 
Then $\big(f_o^*, f_e^* \big)$ is the value of the pair of functions $(f_o, f_e) \in {\cal C}$ that results from this numerical computation.

\medskip

%
%
%

\begin{center}
\bf{\large{APPENDIX B}}
\end{center}

\noindent In this appendix we prove two important theorems on the conditional coverage probability and scaled expected length of the new confidence interval. 

\smallskip

THEOREM B.1.  \textsl{For known 
	$(\sigma_{\varepsilon}, \delta)$, the conditional coverage probability and scaled expected length of 
	$J(f_o, f_e; \sigma_{\varepsilon}, \delta)$ are, for given
	$(f_o, f_e)$, functions of 
	$(\gamma, \delta)$.}

\smallskip

\begin{proof}
Assume that $(\sigma_{\varepsilon}, \delta)$ is known.  Let $h = h(\sigma_{\varepsilon}, \delta)$.
It is straightforward to show that
$P\big( b \in J\big(f_o, f_e; \sigma_{\varepsilon}, \delta\big) \, \big| \, x \big)$,
 the coverage probability of 
$J\big(f_o, f_e; \sigma_{\varepsilon}, \delta\big)$ conditional on $x$, is equal to
\begin{equation*}
P \big( - f_o(h) - f_e(h) \leq g_L \leq - f_o(h) + f_e(h) \, \big | \, x \big),
\end{equation*}
where 
\begin{equation*}
g_L 
= \frac{\widehat{b}_W - b}{\left( \text{Var}(\widehat{b}_W \, | \, x) \right)^{1/2}}
= \frac{\widehat{b}_W - b}{(\sigma_{\varepsilon}^2 \, / \, \SSW)^{1/2}}. 
\end{equation*}
By Theorem 7 of Kabaila \textit{et al} (2017), conditional on $x$, $(g_L, h)$ has a bivariate normal distribution which is determined by $(\gamma, \delta)$.  
It follows that for any given 
$(f_o, f_e) \in {\cal D}$, the coverage probability of 
$J\big(f_o, f_e; \sigma_{\varepsilon}, \delta\big)$ conditional on $x$, is a function of $(\gamma, \delta)$.

The length of $J(f_o, f_e; \sigma_{\varepsilon}, \delta)$ is equal to $2 \, \big(\sigma_{\varepsilon}^2 \, / \, \SSW \big)^{1/2} f_e(h)$.  The length of $L(\sigma_{\varepsilon}, 1 - \alpha)$ is equal to 
$2 \, \big(\sigma_{\varepsilon}^2 \, / \, \SSW \big)^{1/2} z_{1 - \alpha/2}$.
The conditional scaled expected length of 
$J\big(f_o, f_e; \sigma_{\varepsilon}, \delta \big)$ is defined to be
\begin{equation*}
\frac{E\big(\text{length of}\ 
J\big(f_o, f_e; \sigma_{\varepsilon}, \delta \big) \, \big| \, x\big)}
{\text{length of}\ L(\sigma_{\varepsilon}, 1 - \alpha)}
= \dfrac{ E \left( f_e(h) \, | \, x \right)}{z_{1 - \alpha/2}}.
\end{equation*}
By Theorem 7 of Kabaila \textit{et al} (2017), the distribution of $h$, conditional on $x$,  is determined by $(\gamma, \delta)$.  It follows that  the scaled expected length of 
$J\big(f_o, f_e; \sigma_{\varepsilon}, \delta \big)$ is a function of 
$(\gamma, \delta)$. \hfill $\qed$

\end{proof}

\medskip

THEOREM B.2.  \textsl{The conditional coverage probability and scaled expected length of the new confidence interval $CI(\widehat{\sigma}_{\varepsilon}, \widehat{\delta})$ are functions of $(\gamma, \delta)$.}

\smallskip

\begin{proof}
 Define $\widehat{h}$ and $\widehat{g}_L$ to be the values of $h$ and $g_L$, respectively, when $(\sigma_{\varepsilon}, \delta)$ is replaced by $(\widehat{\sigma}_{\varepsilon}, \widehat{\delta})$.  Lemma 1 of Kabaila \textit{et al} (2017) gives explicit expressions for $\widehat{h}$ and $\widehat{g}_L$.
The new confidence interval is
\begin{align}
\label{NewCI}
CI(\widehat{\sigma}_{\varepsilon}, \widehat{\delta}) = 
\left[ 
\widehat{b}_W + \frac{\widehat{\sigma}_{\varepsilon}}{\SSW^{1/2}} \,
f_o^*\big(\widehat{h};\widehat{\delta}, \varphi^*(\widehat{\delta})\big)
\pm \frac{\widehat{\sigma}_{\varepsilon}}{\SSW^{1/2}} \, f_e^*\big(\widehat{h};\widehat{\delta}, \varphi^*(\widehat{\delta})\big) \right]
\end{align}
The coverage probability of this confidence interval, conditional on $x$, is
\begin{equation*}
P \left( -f_o^*\big(\widehat{h};\widehat{\delta}, \varphi^*(\widehat{\delta})\big) - f_e^*\big(\widehat{h};\widehat{\delta}, \varphi^*(\widehat{\delta})\big)
\leq \widehat{g}_L \leq 
-f_o^*\big(\widehat{h};\widehat{\delta}, \varphi^*(\widehat{\delta})\big) + f_e^*\big(\widehat{h};\widehat{\delta}, \varphi^*(\widehat{\delta})\big)\, \Big| \, x \right).
\end{equation*}
Let $\varepsilon_{it}^{\dag} = \varepsilon_{it}/\sigma_{\varepsilon}$ and   $\eta_i^{\dag} = \eta_i/\sigma_{\eta}$.
The $\eta_i^{\dag}$'s and $\varepsilon_{it}^{\dag}$'s are i.i.d. $N(0, 1)$.
By Appendix A.3 of Kabaila \textit{et al} (2017), 
$\widehat{\delta}$ is a function of the $\varepsilon_{it}^{\dag}$'s, $\eta_i^{\dag}$'s, $\delta$ and $x$. Also, by this appendix, 
$\widehat{h}$ and $\widehat{g}_L$ are functions of 
the $\varepsilon_{it}^{\dag}$'s, $\eta_i^{\dag}$'s, $\delta$, $\gamma$ and $x$.
It follows from this that the coverage probability of  the new confidence interval,
conditional on $x$, is
a function of $(\gamma, \delta)$.

Let $c_{\text{min}}$ denote the infimum over $(\gamma, \delta)$ of the coverage probability of the new confidence interval \eqref{NewCI}, conditional on $x$.
As noted by Kabaila \textit{et al} (2007, Section 4), 
for any given $c$, 
$P \big(b \in L(\widehat{\sigma}_{\varepsilon}, c) \, \big| x\,\big)$
does not depend on any unknown parameters. 
Now define $\widetilde{c}$ to be the value of $c$ such that 
$P \big(b \in L(\widehat{\sigma}_{\varepsilon}, c) \, \big| \,x \big) 
= c_{\text{min}}$. 
To a very good approximation, 
$P \big(b \in L(\widehat{\sigma}_{\varepsilon}, c) \, \big| \,x \big) = c$.
For the airline data this approximation is so accurate that with negligible error we may assume that $\widetilde{c} = c_{\text{min}}$ for this data.
The conditional scaled expected length of the new confidence interval
\eqref{NewCI} is defined to be 
$E\big(\text{length of} \ \eqref{NewCI} \, \big| \, x \big)$ divided by $
E\big(\text{length of} \ L(\widehat{\sigma}_{\varepsilon},\widetilde{c}) 
\, \big| \, x  \big)$. To an excellent approximation, this is equal to
\begin{equation}
\label{SEL_expression}
\dfrac{ E \left( (\widehat{\sigma}_{\varepsilon}/ \sigma_{\varepsilon}) \,  f^*_e(\widehat{h}; \widehat{\delta}, \varphi^*( \widehat{\delta})) \, \big| \, x  \right)}{z_{(c_{\text{min}} + 1) / 2} \, E \big( \widehat{\sigma}_{\varepsilon} / \sigma_{\varepsilon} \, | \,  x \big)}.
\end{equation}
By Appendix A.3 of Kabaila \textit{et al} (2007),  $\widehat{\sigma}_{\varepsilon} / \sigma_{\varepsilon}$ and $\widehat{\delta}$  are both
functions of the $\varepsilon_{it}^{\dag}$'s, $\eta_i^{\dag}$'s, $\delta$ and $x$. Also, by this appendix, 
$\widehat{h}$ is a function of 
the $\varepsilon_{it}^{\dag}$'s, $\eta_i^{\dag}$'s, $\delta$, $\gamma$ and $x$.
Hence \eqref{SEL_expression} is
a function of $(\gamma, \delta)$.
 \hfill $\qed$

\end{proof}

\medskip

\begin{center}
\bf{\large{APPENDIX C}}
\end{center}

\noindent \textbf{Estimating the coverage probability of the new confidence interval by simulation}

\medskip

\noindent In this section we describe the estimation of the coverage probability of the new confidence interval.
Consider $M$ independent simulation runs, indexed by $k$.  On the $k$'th simulation run, we do the following.  Generate observations of $\eta^{\dag} = (\eta_1^{\dag}, \dots, \eta_N^{\dag})$ and $\varepsilon^{\dag} = (\varepsilon_{11}^{\dag}, \dots, \varepsilon_{1T}^{\dag}, \dots, \varepsilon_{N1}^{\dag}, \dots, \varepsilon_{NT}^{\dag})$ where the $\eta_i^{\dag}$'s and $\varepsilon_{it}^{\dag}$'s are i.i.d. $N(0, 1)$.
Compute $\widehat{\delta}_k$, $\widehat{h}_k$ and $\widehat{g}_{L, k}$, the values of $\widehat{\delta}$, $\widehat{h}$ and $\widehat{g}_L$ for this simulation run, respectively.  
Find the functions $f_o^*\big(\, \cdot \, ;\widehat{\delta}_k, \varphi^*(\widehat{\delta}_k)\big)$ and $f_e^*\big(\, \cdot \,; \widehat{\delta}_k, \varphi^*(\widehat{\delta}_k)\big)$ for this simulation run by the method described in the second and third paragraphs of Section \ref{sim_method}.  
Let $f^*_{o, k}(\, \cdot \,) = f^*_o\big( \, \cdot \, ; \widehat{\delta}_k, \varphi^*(\widehat{\delta}_k) \big)$ and $f^*_{e, k}(\, \cdot \,) = f^*_e\big( \, \cdot \, ; \widehat{\delta}_k, \varphi^*(\widehat{\delta}_k) \big)$.
Define the function
\begin{equation*}
{\cal I}({\cal A}) = \begin{cases} 1 & \text{if ${\cal A}$ occurs} \\ 0 & \text{if ${\cal A}^c$ occurs}, \end{cases}
\end{equation*}
where ${\cal A}$ is an arbitrary event.
The simulation based estimator of the coverage probability that we use is
\begin{equation*}
\widehat{\textsc{CP}} = \dfrac{1}{M} \sum_{k=1}^M {\cal I} \left( - f^*_{o, k}\big(\widehat{h}_k\big) - f^*_{e, k}\big(\widehat{h}_k\big) \leq \widehat{g}_{L, k} \leq - f^*_{o, k} \big(\widehat{h}_k\big) + f^*_{e, k}\big(\widehat{h}_k\big) \right)
\end{equation*}
The variance of $\widehat{\textsc{CP}}$ can be estimated using properties of a binomial proportion.

\medskip

\noindent \textbf{Estimating the confidence coefficient of the new confidence interval by simulation}

\medskip


\noindent In this section we describe the estimation of the confidence coefficient $c_{\text{min}}$ of the new confidence interval.
We specify a grid of $\delta$ values and a grid of $\gamma$ values.
For each of these $\delta$ values we find the coverage probability, minimized over $\gamma$, as follows.  Using $M_1$ independent simulation runs, we estimate the coverage probability at each of these $\gamma$ values.  Store the $\gamma$ values that lead to the three smallest estimated coverage probabilities.  For each of these three $\gamma$ values, estimate the coverage probability again using $M_2$ independent simulation runs, where $M_2$ is substantially larger than $M_1$.  Store the value of $\gamma$ which leads to the smallest estimated coverage probability.
Once this is done for each $\delta$, we obtain the $\delta$ and $\gamma$ values that minimize the estimated coverage probability over the grid of values of $\delta$ and $\gamma$.  
Denote these values by $\gamma^*$ and $\delta^*$, respectively.  
To ensure this estimate is not biased downwards, we estimate the coverage probability once more using $M_3$ simulations for true parameter values $\delta^*$ and $\gamma^*$, where $M_3$ is substantially larger than $M_2$.  
The resulting coverage probability is our estimate of $c_{\text{min}}$.
This search for the minimum coverage probability is similar to that described in Section 3.1 of Kabaila and Leeb (2006).

 For the airfare data, the grid of $\gamma$ values is $-200, -190, \dots, 190, 200$ and the grid of $\delta$ values is $0, 2.5, \dots, 15, 20, 30, 50, 80$.  Note that low values of $\delta$ correspond to large values of $\rho(\delta)$, given by \eqref{eqn_rho}.  We used $M_1 = 100,000$, $M_2 = 1,000,000$ and $M_3 = 4,000,000$.  The resulting estimate of $c_{\text{min}}$ was 0.9493.  The standard error of this estimate of the coverage probability at $\delta^*$ and $\gamma^*$ is $1.1 \times 10^{-4}$.  

\medskip

\noindent \textbf{Estimation of the scaled expected length of the new confidence interval by simulation}

\medskip


\noindent We estimate the expected values in the numerator and denominator of the expression for the scaled expected length, given by equation \eqref{SEL_expression}, by separate sets of $M$ independent simulation runs.  
We begin by describing the estimation of the numerator, $E \Big( (\widehat{\sigma}_{\varepsilon} / \sigma_{\varepsilon}) \, f_e^*(\widehat{h}; \widehat{\delta}, \varphi^*(\widehat{\delta})) \, | \, x \Big)$.
On the $k$'th simulation run, we do the following.  Generate observations of $\eta^{\dag} = (\eta_1^{\dag}, \dots, \eta_N^{\dag})$ and $\varepsilon^{\dag} = (\varepsilon_{11}^{\dag}, \dots, \varepsilon_{1T}^{\dag}, \dots, \varepsilon_{N1}^{\dag}, \dots, \varepsilon_{NT}^{\dag})$, where the $\eta_i^{\dag}$'s and $\varepsilon_{it}^{\dag}$'s are i.i.d. $N(0, 1)$.
Compute $(\widehat{\sigma}_{\varepsilon} / \sigma_{\varepsilon})_k$, $\widehat{\delta}_k$ and $\widehat{h}_k$, the values of $(\widehat{\sigma}_{\varepsilon} / \sigma_{\varepsilon})$, $\widehat{\delta}$ and $\widehat{h}$, respectively, for this simulation run.  
Find the function $f_e^*\big(\, \cdot \,; \widehat{\delta}_k, \varphi^*(\widehat{\delta}_k)\big)$ for this simulation run by the method described in the second and third paragraphs of Section \ref{sim_method}.  
Let $f^*_{e, k}(\, \cdot \,) = f^*_e\big( \, \cdot \, ; \widehat{\delta}_k, \varphi^*(\widehat{\delta}_k) \big)$.
The simulation based estimator of 
$E \Big( (\widehat{\sigma}_{\varepsilon} / \sigma_{\varepsilon}) \, f_e^*(\widehat{h}; \widehat{\delta},\varphi^*(\widehat{\delta})) \, | \, x \Big)$ that we use is 
\begin{equation*}
\textsc{NUM} = \dfrac{1}{M} \sum_{k=1}^M (\widehat{\sigma}_{\varepsilon} / \sigma_{\varepsilon})_k \, f^*_{e, k}(\widehat{h}_k).
\end{equation*}

We now describe the estimation of the term in the denominator, $E \big( (\widehat{\sigma}_{\varepsilon} / \sigma_{\varepsilon}) \, | \, x \big)$, of the right-hand side of \eqref{SEL_expression}.  Consider another set of $M$ independent simulation runs.  On the $k$'th simulation run we do the following.  Generate observations of $\eta^{\dag}$ and $\varepsilon^{\dag}$. 
Let $(\widehat{\sigma}_{\varepsilon} / \sigma_{\varepsilon})_k$ denote the value of $\widehat{\sigma}_{\varepsilon} / \sigma_{\varepsilon}$ for this simulation run.  The simulation based estimator of $E \big( (\widehat{\sigma}_{\varepsilon} / \sigma_{\varepsilon}) \, | \, x \big)$ that we use is 
\begin{equation*}
\textsc{TERM} = \dfrac{1}{M} \sum_{k=1}^M (\widehat{\sigma}_{\varepsilon} / \sigma_{\varepsilon})_k.
\end{equation*}

Overall, the simulation based estimator of the scaled expected length \eqref{SEL_expression} is 
\begin{equation*}
 \dfrac{\textsc{NUM}}{z_{(c_{\text{min}} + 1) / 2} \, \textsc{TERM}},
\end{equation*}
where $c_{\text{min}}$ is estimated using the simulation method described in the previous subsection.


\end{document}